\documentclass[acmsmall,screen,authorversion,nonacm]{acmart}
\usepackage{tabulary}
\usepackage{hyperref}

\usepackage{fancyhdr}
\AtBeginDocument{%
    \addtolength{\footskip}{2.0\baselineskip}%
    \fancyfoot[R]{\textit{Accepted at CSCW 2024.}}%
}

\AtBeginDocument{%
  }

\setcopyright{none}
\acmYear{2024} \acmPrice{15.00}\acmDOI{XXXXXX}

\begin{document}


\title{Understanding Practices around Computational News Discovery Tools in the Domain of Science Journalism}

\author{Sachita Nishal}
\email{nishal@u.northwestern.edu}
\affiliation{%
  \institution{Northwestern University}
%  \city{Evanston, Illinois}
  \country{USA}
  }

\author{Jasmine Sinchai}
\email{jasminesinchai2025@u.northwestern.edu}
\affiliation{%
  \institution{Northwestern University}
%  \city{Evanston, Illinois}
  \country{USA}
  }
  
\author{Nicholas Diakopoulos}
\email{nad@northwestern.edu}
\affiliation{%
  \institution{Northwestern University}
%  \city{Evanston, Illinois}
  \country{USA}
  }

\begin{abstract}

Science and technology journalists today face challenges in finding newsworthy leads due to increased workloads, reduced resources, and expanding scientific publishing ecosystems. Given this context, we explore computational methods to aid these journalists' news discovery in terms of time-efficiency and agency. In particular, we prototyped three computational information subsidies into an interactive tool that we used as a probe to better understand how such a tool may offer utility or more broadly shape the practices of professional science journalists. Our findings highlight central considerations around science journalists' agency, context, and responsibilities that such tools can influence and could  account for in design. Based on this, we suggest design opportunities for greater and longer-term user agency; incorporating contextual, personal and collaborative notions of newsworthiness; and leveraging flexible interfaces and generative models. Overall, our findings contribute a richer view of the sociotechnical system around computational news discovery tools, and suggest ways to improve such tools to better support the practices of science journalists. 
\end{abstract}




\maketitle

\section{Introduction}

Science journalism is a complex endeavor with wide-ranging objectives like chronicling new breakthroughs, explaining the scientific process, and contextualizing broader problems \cite{angler_science_2017}, with potential impacts on people's personal decision-making processes as well as on scientific policy and legislation \cite{nisbetNeedKnowledgeBasedJournalism2015, shoemakerJournalistsGatekeepers2009}. The science journalist's first step in this crucial work is the task of \textit{news discovery}, when they first encounter a potentially newsworthy event or issue (i.e. a \textit{lead}) \cite{mcmanus_market-driven_1994}. Over the past few decades, the intensity of this task has been amplified by a host of institutional and social changes. These shifts have compelled journalists to cover multi-faceted topics of greater societal import by adopting different roles and formats \cite{blumScienceJournalismGrows2021, fahy2011science}, while simultaneously monitoring the skyrocketing volume of articles from scientific journals, preprint servers, conference proceedings, corporate blogs, and so on \cite{bornmann2021growth}. This creates the risk of lower quality in their reporting, and even professional burnout \cite{andersonViewTrenchesInterviews2023, powellWhatScienceJournalism2015b}. Science journalists thus actively seek to increase their efficiency in uncovering, understanding, and critiquing scientific information \cite{andersonViewTrenchesInterviews2023}. 

One efficiency mechanism that science journalists leverage is using \textit{information subsidies}, which are easily digestible documents that simplify newsworthy information and technical jargon from scientific articles \cite{gandy_information_1980}. For instance, press releases are a familiar type of information subsidy that supports the science communication process when available \cite{maclaughlin2018predicting}, though they are not without their own issues in how they may bias or shape coverage \cite{sumner_exaggerations_2016, vogler_growing_2020}. In this work, we adopt a computational approach to create more widely accessible information subsidies, with the long-term goal of reducing the effort and increasing the stock of public-interest science journalism \cite{Hamilton:2016ur}. Such a \textit{computational news discovery} (CND) approach has been defined as "the use of algorithms to orient editorial attention to potentially newsworthy events or information prior to publication." \cite{diakopoulosComputationalNewsDiscovery2020}. The objective of CND is to reduce the time cost of identifying newsworthy leads and alerting journalists to their existence. Extant CND tools monitor various information sources to suggest leads, provide context, and sometimes even rank them by "newsworthiness" \cite{diakopoulos2021towards, liu2017reuters, magnusson2016finding, petridisAngleKindlingSupportingJournalistic2023}. These systems offer a valuable starting point so that journalists can then exercise their expertise and creativity for selecting what becomes news. 

In order to design CND tools that align with journalists’ practices, it is important to understand their perspectives and existing workflows - particularly in specific domains like science journalism. Thus, in this study, we prototyped \textit{computational information subsidies} and displayed them in an interactive tool, that we then used as a stimulus in semi-structured interviews with professional science journalists, to understand user-centered perspectives and practices around CND. We were broadly motivated by questions such as: How do journalists engage with and question different computational subsidies for evaluating "newsworthiness"? How do they envision the affordances of these subsidies in their workflows? How do these subsidies interact with broader constraints and tensions within science journalism? To explore these questions we prototyped three subsidies for scientific articles: (1) a \textit{newsworthiness score} based on an article abstract, (2) an \textit{outlet relevance score} to measure the suitability of the article for different news outlets, and (3) a set of \textit{news angles} that suggest different approaches to write a story about the article. We displayed these subsidies and article metadata in the tool's interactive user interface that journalists browsed during their interviews.

Through an inductive thematic analysis of these interviews, we uncover several user benefits, expectations, and concerns for CND tools in the science journalism domain. We identify common dimensions of journalists' agency that CND could support, such as by helping strategize lead exploration, increasing source diversity, and enabling scientific understanding. This can counterbalance the complexity of the technical information landscape science journalists work in. We also find tangible, context-specific aspects of journalistic practice that these subsidies can accommodate for added flexibility, such as audience focus, labor status, writing format, and experience level. And we identify overarching tensions in news production that can be influenced by these subsidies, such as competition across news outlets, balance of editorial and personal interest in newsrooms, and societal patterns in concentration of attention. 

In discussing these findings, we further elaborate how they build on prior understandings of science journalism practices and complexities of newsworthiness evaluation. We also suggest several design possibilities to support journalists' agency, incorporate personal and contextual needs, and develop new features for newsworthiness evaluation. Overall, our work not only highlights the benefits and impacts of CND for science journalists, but it does so by deepening the understanding of domain-specific workflows, interpersonal dynamics, and professional responsibilities that CND tools would be enmeshed within if deployed. This work therefore contributes to a richer view of the sociotechnical system that emerges around computational news discovery tools, and paves the way for improved tools that are sensitive to the nuanced needs and practices of science journalists. 

\section{Related Work}

In this section, we highlight current challenges to news discovery faced by science journalists. This puts forth a set of informational and interactional needs for our users, which help motivate our computational approach and evaluation study. Following this, we highlight the specific computational subsidies we designed to address these needs, the prior work they build on, and what we hope to learn from their use in our interview study about the practices, concerns and expectations of science journalists with regards to CND.

\subsection{Science Journalists' Challenges for News Discovery}
\label{subsection-sciencejournalism}

Science journalists discover the latest scientific breakthroughs through various channels: they refer to press releases and e-mail alerts from universities and journals, recent issues of major journals, preprint servers for different disciplines, major government websites, scientists' and corporate blogs, tips from existing sources, and even listservs and social media \cite{blum_field_2006, fleerackersScienceMotionQualitative2022, dunwoody2014science, oranskyJournalsPeerReview2022, maiden_designing_2020}. Keeping up with the latest research provides them with exciting story ideas, and also develops their news judgment \cite{blum_field_2006}. This process of monitoring information for potentially newsworthy leads is referred to as \textit{news discovery} \cite{mcmanus_market-driven_1994, reich_process_2006}. It can result in different types of stories, such as a \textit{study story} about a particular research article, or a \textit{feature story} about a broader trend.
%As veteran science journalist Phillip Yam puts it: "The more you know what’s going on, the better you will be at recognizing a good story when it comes along." \cite{blum_field_2006} (p. 7). 

In recent years, news discovery has become more challenging for science journalists due to various factors:

\begin{itemize}
    \item \textbf{Pluralizing Responsibilities}: Science journalists today not only report on scientific advancements but also serve as watchdogs, curators, explainers, and public intellectuals \cite{fahy2011science, blumScienceJournalismGrows2021}. Their work influences policy-making, science education, and agenda-setting. To fulfill these roles and enable multiple perspectives in their work, they need to monitor publications, retractions, discourse, and practices in science \cite{rensbergerScienceJournalismToo2009, oranskyJournalsPeerReview2022}, all the while sifting through a fire-hose of information, misinformation, and commentary \cite{russellScienceJournalismGoes2009}. 

    \item \textbf{Shrinking Resources}: Due to reduction in science news desks, reporting budgets, and freelance clients, science journalists are facing increased workloads \cite{dunwoody2014science, bajak2016economics, nijhuis_freelancing_2015, powellWhatScienceJournalism2015a, vergano_gig_2014, scheufele2013communicating}. They rely more on \textit{information subsidies} \cite{gandy_information_1980} from science institutions (e.g. press releases or PRs) for efficiently sourcing jargon-free leads \cite{brumfielScienceJournalismSupplanting2009, fahy2011science}. Empirical studies show that press releases significantly influence news coverage \cite{de_semir_press_1998, machill_influence_2006, maclaughlin2018predicting}. However, PRs are only available for select high-prestige institutions and journals, create an information overload \cite{smithBreakingNewsDesign2018}, and relying too heavily on them can perpetuate a mode of "passive newsmaking" \cite{comfort_building_2022} and undermine journalistic principles like objectivity, accuracy, and independence \cite{brumfielScienceJournalismSupplanting2009, sumner_exaggerations_2016, vogler_growing_2020}.

    \item \textbf{Expanding Scientific Ecosystem}: The volume of journalists' scientific sources has grown exponentially since the 1950s \cite{bornmann2021growth, fortunato_science_2018}, increasing time pressures for news discovery. One such source is preprints, which are preliminary versions of papers available freely before formal publication. Preprint sharing is especially common in computer science and its various sub-domains like ML, HCI, and so on. They provide journalists timely and accessible findings for public dissemination across various fields \cite{fleerackersScienceMotionQualitative2022, fraser_evolving_2021, hutsonArtificialIntelligence2022, priorScienceWritersShould2022, guptaSocialSciences2022, maiden_designing_2020}. However, journalists need to approach preprints with higher skepticism than other, peer-reviewed work \cite{priorScienceWritersShould2022, fleerackersScienceMotionQualitative2022, maiden_designing_2020}, while at the same time they are not typically accompanied by PRs to enable their sense-making. 
\end{itemize}

In light of these shifts toward intensified work and strained attention, we explore how CND tools could help science journalists discover newsworthy leads in a way that is efficient and reliable. As these shifts have shown, CND tools for science journalism must support discovery and sense-making of highly technical work, to enable deliberation from different perspectives and angles, and even in the absence of press releases. They must also incorporate central editorial and journalistic values. 

Newsrooms increasingly utilize AI/ML-based tools to support journalistic needs, including and beyond CND \cite{beckettNewPowersNew2019, rinehartArtificialIntelligenceLocal2022, jonesAIEverywhereNowhere2022, liu2017reuters, diakopoulosGeneratingLocationBasedNews2020}. Recent advancements in AI have elicited deeper reflection and transparency from editors around the role such tools can and should play in newsrooms, and how they may impact or incorporate editorial values \cite{nunezLetterEditorHow2023, vinerGuardianApproachGenerative2023}. In anticipation of and in response to such deployments, there has also been deeper work in the HCI community about the utility, efficiency and perception of these tools for a wide variety of tasks, such as news discovery \cite{diakopoulos2021towards}, information verification \cite{tolmie_supporting_2017}, story framing \cite{petridisAngleKindlingSupportingJournalistic2023}, and article generation \cite{oh_understanding_2020}.

However, studies find that the actual \textit{adoption} of computational tools by journalists varies significantly due to lower levels of trust, uncertainty about AI/ML capabilities, and the logistical difficulty of integrating them into actual journalistic workflows \cite{roperGaugingGenerativeAI2023, dehaanInvisibleFriendFoe2022, gutierrezlopezQuestionDesignStrategies2022, beckettNewPowersNew2019, moranRobotsNewsNewsrooms2022}. Further, these studies show that journalists constantly grapple with how automated tools impact their identities, daily tasks, and the democratic role of journalism \cite{moranRobotsNewsNewsrooms2022}. This too can be difficult when journalists are unsure of the nature of these systems or about how to best integrate them into their practices. 

One way to support this kind of trust, adaptation, and deliberation is via the means of AI literacy programs to educate journalists about the mechanisms and caveats of the tools they use \cite{deuzeImaginationAlgorithmsNews2022, caiMotivationsGoalsPathways2023}. However, it is also important to support these activities while users are engaging with said tools. Recent work has demonstrated that this requires intelligible, contextually-specific interfaces that visibly align with editorial and ethical requirements, and which are consistently under users' control \cite{jonesAIEverywhereNowhere2022, abdul_trends_2018}. However, there is a relative paucity of work for building and studying such interfaces in the context of science journalism, as compared to other domains \cite{maidenAutomatingScienceJournalism2023}.

In this work, we specifically set out to probe what intelligible, contextual, and editorially-aligned interfaces can look like in the context of CND for science journalism. To this end, we prototype different computational subsidies based on the journalistic needs articulated above, and use them as a way to elicit science journalists' perspectives on how CND tools can integrate with and shape their practices, values and industry, which can help develop effective tools in the future. We focus our evaluation on discovering news from pre-prints, specifically the arXiv Computer Science (CS) pre-print server\footnote{https://arxiv.org}, since pre-prints typically pose heightened challenges to robust journalistic decision-making, often lack traditional information subsidies, and also offer readily available basic metadata for the subsidies. 

\subsection{Motivating Subsidies for Science News Discovery}
\label{subsection-cnd}

We leverage prior research to design three computational subsidies, namely \textit{newsworthiness scores}, \textit{outlet relevance scores}, and \textit{news angles}, aiming to explore the potential benefits and impacts of CND for science journalists' practices. These subsidies rely on statistical approaches to textual similarity measurement, feature-based prediction, and text generation models. Our contribution lies in applying these established techniques to create composite application-level functionality and explore its implications for science journalists, who have unique skills, needs, and constraints. The following section provides an overview of the relevant research that informs this design and evaluation.

CND entails the development of automated tools that direct journalists towards newsworthy information at the initial stage of news discovery \cite{diakopoulosComputationalNewsDiscovery2020}. To vet a lead for newsworthiness, journalists traditionally rely on several contextual, social and institutional factors. These factors are known as \textit{news values}  \cite{ harcup_what_2001, badenschier_2012}, and include qualities of the lead such as its novelty, controversy, impact, and so on. Logistical and commercial factors such as the potential audience interest in a story, the reporting budget, and the diversity of the weekly newspaper spread can also influence whether a story makes it into the press \cite{allern_journalistic_2002}. CND seeks to leverage computing to evaluate potential news leads according to these factors. 

Extant systems for CND largely monitor social media feeds and structured documents as they aim to surface recent or even real-time events for further journalistic scrutiny \cite{diakopoulosComputationalNewsDiscovery2020, diakopoulosGeneratingLocationBasedNews2020, diakopoulos2021towards, liu2017reuters, magnusson2016finding, schwartz_editorial_2015}. These systems automatically evaluate such leads for various news values, and use those to enable judgments of newsworthiness. The sources monitored and the news values chosen often depend on the specific journalistic domain. For instance, the Reuters Tracer is designed to surface breaking news from Twitter feeds, and as such relies on evaluation of events like man-made and natural disasters according to news values such as \textit{magnitude}, \textit{human impact}, and \textit{financial impact} \cite{liu2017reuters}. Similarly, the  AlgorithmTips system surfaces newsworthy leads from government documents pertaining to algorithmic decision-making systems, using news values like \textit{novelty}, \textit{controversy}, \textit{organizational agenda}, and so on \cite{diakopoulos2021towards}. In this vein, the first information subsidy we design is a \textit{newsworthiness score} for each abstract, which is generated based on prominent news values in science journalism such as \textit{actuality}, \textit{controversy}, \textit{magnitude of impact}, and \textit{valence of impact} \cite{badenschier_2012}. 

Commercial news values pertaining to news production and distribution also influence the likelihood of a story being published and must often be negotiated with other public interest responsibilities \cite{petre_all_2021}. We provide support for these by developing and evaluating a second information subsidy: the \textit{outlet relevance score}. This score can support journalists as they consider whether a story might be suitable to pitch at a chosen news outlet \cite{diakopoulosComputationalNewsDiscovery2020}, based on the stories it has published in the past, which can indicate what an outlet's readers might prefer. For instance, \textit{general news outlets} cater to a broad audience and provide simplified coverage of scientific topics. Meanwhile, \textit{popular science outlets} engage a broad readership, presenting scientific information in an accessible and entertaining way. \textit{Trade outlets} will instead focus on specialized scientific fields, providing in-depth coverage and targeting professionals within those domains. Preferences for themes and topics of stories differ across and within these types of outlets as well. While others have argued that commercial news values are typically internalized within the decision-making frameworks of news production \cite{allern_journalistic_2002}, we leverage publicly available data to develop a proxy that can be applied at scale without requiring access to internal information.  

Tools have also been designed to support journalists’ creativity as they summarize jargon-heavy information for laypersons \cite{dangovski_we_2021}, look for inspiration to begin writing \cite{gero_design_2022, petridisAngleKindlingSupportingJournalistic2023}, and brainstorm angles for stories based on related news \cite{maiden_making_2018}. We drew from this literature to motivate and design an information subsidy around \textit{news angles} generated based on the abstract of the research article. News angles identify pivotal elements and narratives within a scientific abstract \cite{temmerman_news_2021}, and we present them to help journalists consider multiple frames for a lead. For instance, a scientific finding might be framed in terms of the its impacted populations, economic outcomes, or controversial results. Each news angle thus helps to select, shape, and focus information in ways that can lead to different potential stories. The presence of news angles not only provides more insight to a journalist as they evaluate a lead's news potential, but it can also kick-start the \textit{news gathering} process \cite{reich_process_2006} by helping them conceptualize further avenues that could be explored or by considering specific kinds of outlets that might be interested in a given story.

Put together, these subsidies allow for the evaluation of central journalistic and editorial values (i.e. news values and outlet suitability) for a body of technical work that is not typically accompanied by PR (i.e. preprints), along with a review of multiple perspectives (i.e. news angles) that might be adopted toward a story. 

Subsidies focusing on newsworthiness-based ranking and news angle generation for newsworthy documents have been evaluated in recent studies involving journalists, although their perspectives differ from the approach we undertake in this research. One approach utilized crowdsourcing to rank potential leads for investigative journalists, and it analyzed how users conceptualize "newsworthiness" and make plans to turn leads into stories using the tool \cite{diakopoulos2021towards}. Our implementation of newsworthiness and outlet fit metrics draws from their ranking-based approach, but relies entirely on machine learning, allowing us to raise questions specifically around the scale, trust, and transparency of automated tools for CND. Another study highlighted the use of large language models to generate news angles from political press releases, and the tool's ability to help journalists engage with this work and ideate on it for different types of stories \cite{petridisAngleKindlingSupportingJournalistic2023}. In our work, we focus on enabling ideation specifically within the domain of science journalism, taking into account the domain-specific considerations discussed in the previous section, and focusing specifically on qualitative perceptions of the feature. We also expect that our broader alignment with these works positions us to draw from established ideas in CND, while letting us uncover the unique opportunities and tensions that arise for our set of users and our specific domain.

\section{Designing Computational Subsidies for News Discovery}
\label{section-designstuff}

This section describes the development and presentation of the three different information subsidies. We synthesize these subsidies into a coherent interactive user interface (UI), as a way to ground the interviews we undertook with practicing science journalists. We make no claim that our implementations of these subsidies are the only or best way to do so, however, we provide enough detail below so that the findings from our interview study can be interpreted accurately with respect to what users experienced in the interface. Wherever we refer to the "tool", we are referring to this interactive UI incorporating the various subsidies and article metadata.

% WHAT WE PRESENT - EDITORIAL GOALS
We broadly draw from scholarship guiding and evaluating the design of AI-infused tools i.e. having "features harnessing AI capabilities that are directly exposed to the end user" \cite{amershiGuidelinesHumanAIInteraction2019}. Our main objective was to center the presentation of information that aligns with users' contextual needs \cite{amershiGuidelinesHumanAIInteraction2019,diakopoulosComputationalNewsDiscovery2020}, which in this case are journalistic needs for newsworthiness evaluation \cite{badenschier_2012, allern_journalistic_2002}. The previous section has highlighted how we accounted for these when selecting computational subsidies.
% HOW WE PRESENT IT - CONFIGURABILITY/AGENCY
Our secondary goal was to strike a balance between users' agency and the computational subsidies \cite{greenAlgorithmintheLoopDecisionMaking2020, heer2019agency} i.e. clearly position these subsidies in a way to enable journalists' independent decision-making and exercise of domain expertise, and not replace them or automate this crucial task. This allows us to align with central journalistic values such as independence and objectivity, which are crucial for the success of AI tools \cite{komatsuAIShouldEmbody2020, diakopoulosComputationalNewsDiscovery2020}. The tool's UI followed from prior work and allowed end-users to rank and filter leads based on computed scores, instead of providing any outright recommendations \cite{diakopoulosComputationalNewsDiscovery2020, diakopoulos2021towards}. This put the onus of both invoking and ignoring the system's outputs in the hands of journalists \cite{amershiGuidelinesHumanAIInteraction2019}. The UI also provided article metadata as well as an outgoing link to its webpage to enable the independent verification of the displayed scores and news angles
\cite{gutierrezlopezQuestionDesignStrategies2022}.
% HOW WE FRAME IT - TRANSPARENCY
Finally, we aimed to provide transparency disclosures to end-users about the algorithmic components that made up this system, aiming to foster a general understanding and trust \cite{diakopoulosComputationalNewsDiscovery2020, weld2019challenge, gutierrezlopezQuestionDesignStrategies2022}. The disclosures we provided consisted of textual descriptions and references for the technology underlying individual information subsidies (e.g. GPT, similarity metrics).

%These disclosures along with the interview questions were intended to provoke discussions about deeper information regarding training data, model details, and evaluation methods that were needed to enable trust, without creating further information overload. 

\begin{figure}[t]
  \centering
  \captionsetup{justification=centering,margin=0.15\linewidth}
  \includegraphics[width={0.8\linewidth}]{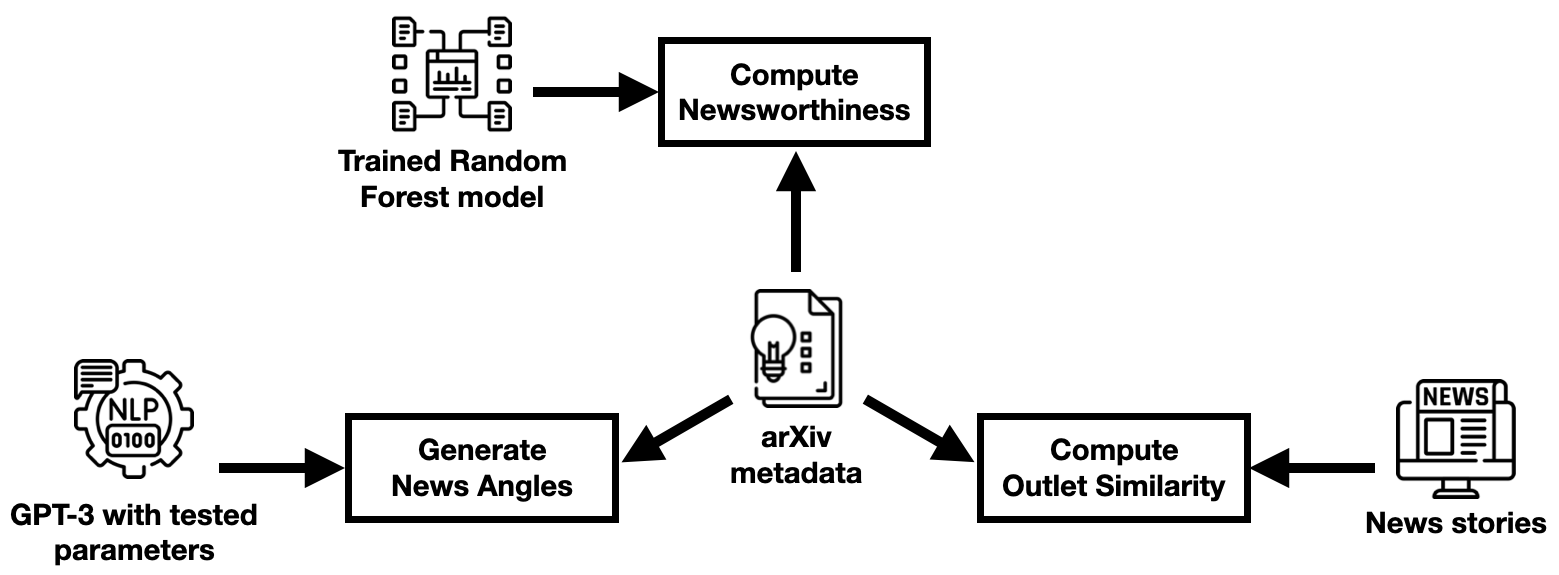}
  \caption{Overview of data and models used to generate the three prototype computational subsidies that were displayed in the UI}
  \Description[Pipeline of data processing and the algorithms used to generate computational subsidies]{A flow chart showing how newsworthiness, outlet fit, and news angles were computed.}
  \label{fig:flowchart_icons}
\end{figure}

Via the interface, we sought to understand if and how journalists would use these features, if they had other contextual or professional needs and concerns that could drive future designs, and how they believed such tools could integrate with and shape their workflows and newsrooms. To generate leads for review, we periodically collected articles from the arXiv CS preprint server, and augmented them with computationally-generated features that could enable newsworthiness evaluation (Section \ref{subsection-datacollection}). We then operationalized two information subsidies for ranking: \textit{newsworthiness score} (Section \ref{subsection-methodsnewsworthiness}) and the \textit{outlet relevance score} (Section \ref{subsection-methodsoutlet}). We also generated \textit{news angles} on articles using GPT-3 (Section \ref{subsection-methodsangles}). Figure \ref{fig:flowchart_icons} provides an overview of the data and models for computing each subsidy. Finally, we presented relevant article metadata and the computed subsidies in the UI (Section \ref{subsection-uidesign}). 

\subsection{Collecting arXiv Articles}
\label{subsection-datacollection}

We first collected and processed articles from the arXiv preprint server for input to the system. We chose the Computer Science (CS) subject area of the arXiv server to extract articles that could serve as leads for science and technology journalists. We collected the title, abstract, URL, primary category, other listed categories, the date of publication, and the full-texts of all the relevant articles\footnote{Using the arxivscraper \cite{sadjadi_arxivscraper_2017} and pdfx \cite{hager_pdfx_2021} libraries}. We also computed features for later use in calculating newsworthiness. For instance, we derived a readability score using the science De-Jargonizer taxonomy \cite{rakedzon_automatic_2017}, which assigns an easy/medium/hard score to scientific terms based on their frequency of occurrence in articles published by the BBC. Our final dataset contains arXiv CS articles and their associated metadata starting with a publication date of January 1, 2017 through Sept 9, 2022. We partitioned this dataset into two subsets, using the data from 2017-2021 for all model training, piloting, experiments as needed ($\sim$85,500 articles), and the more recent data from 2022 for our user study ($\sim$18,500 articles).

\subsection {Computing Newsworthiness}
\label{subsection-methodsnewsworthiness}

We operationalized and calculated a \textit{newsworthiness score} for each article in our dataset using our prior work \cite{nishalCrowdRatingsPredictive2022}, where we refer the reader for an explanation of the method in full detail. Briefly, the method we used leverages crowdsourced evaluations of the following news values observed in the domain of science journalism \cite{badenschier_2012}: \textit{actuality}, \textit{controversy}, \textit{impact magnitude}, and \textit{impact valence}. We aggregated these quantitative evaluations into an overall \textit{newsworthiness score} for each arXiv CS abstract, which ranges from a scale of 0 to 100, with higher values indicating a higher newsworthiness. This score was also validated against evaluations of overall newsworthiness provided by two experts, each respectively with 3 and 4 years of experience in science reporting. The Spearman correlations between aggregated expert and crowdsourced \textit{newsworthiness scores} were calculated over a small validation set of 55 abstracts, sampled in a random and stratified manner from arXiv CS and its subcategories, and indicated a moderate level of agreement (r(55) = 0.379, p = 0.004). We then trained a Random Forest model to predict an abstract's \textit{newsworthiness score} using various textual features. We ranked abstracts in descending order of their \textit{newsworthiness score}, and computed precision@K metrics against expert evaluations. Precision@K measures the proportion of the top-K items ranked by the \textit{newsworthiness score} that were actually newsworthy according to experts as well; the performance was P@10 = 0.8 and P@15 = 0.67. Although there is room for improvement, we consider this adequate performance for the needs of supporting a viable prototype experience for users. The trained prediction model was used to output a \textit{newsworthiness score} for each arXiv article in the dataset for our user study. 

\subsection {Computing Audience Fit for Stories}
\label{subsection-methodsoutlet}

We also computed an \textit{outlet relevance score} to support the evaluation of audience fit for a story based off of a given arXiv article, and for specific news outlets that journalists write and pitch for. For instance, technically-oriented articles can be more suited to trade-focused publications, whereas more directly society-facing research projects lend themselves well to a general news audience. This \textit{outlet relevance score} evaluates the similarity of arXiv articles to the past repertoire of stories from individual outlets. Conceptually speaking, it captures whether an article can reasonably fit in with \textit{relevant} prior coverage at a given outlet, i.e. whether the outlet has prior stories that touch on similar themes or topics. For instance, an arXiv CS article revolving around data privacy could be similar to several articles at MIT Technology Review pertaining to this theme.  For it to be suitable for this publication, the article need not exhibit a high similarity to all their past coverage, but rather it only needs to be relevant to the data privacy niche. 

To calculate these \textit{outlet relevance scores}, we created a list of the top fifty outlets that had most frequently included news coverage of arXiv articles from 2017-2021\footnote{Using the Altmetric API: https://www.altmetric.com}. We filtered this list to exclude websites that publish in non-English languages, that display press releases primarily, and that often publish cloned articles from other prominent outlets. We then collected all news items from the remaining sources\footnote{Using the news-please tool \cite{hamborg_news-please_2017}}. Appendix \ref{appendixA} lists the final set of outlets. 
 
For every news outlet, each news item was cleaned to remove advertisements in the text, and then converted to SentenceBERT embeddings \cite{reimers_sentence-bert_2019}. We then calculated the cosine similarity scores between the SentenceBERT embeddings of each news item and each arXiv CS abstract. Cosine similarity of neural embeddings is widely used as a proxy for the semantic similarity of documents \cite{chandrasekaranEvolutionSemanticSimilarity2022, mathurPuttingEvaluationContext2019, Zhang2020BERTScore}, and has also been used in prior work within the use-case of journalism \cite{wangJournalisticSourceDiscovery2021}. A higher similarity score implies greater semantic similarity for an arXiv article to a news item. For each arXiv article, we selected the top decile of cosine similarities i.e. similarities to the top 10\% most semantically similar news items, and averaged their cosine similarities. The \textit{outlet relevance score} for an article thus reflects how similar it is, on average, to the set of news items that it is most similar to (i.e. its potential niche), for a given outlet. To reiterate, this can help determine the thematic or topical niche that an abstract could be relevant to at a chosen news outlet. In practice we expect there to be a "sweet spot" of outlet relevance which is neither too similar (e.g. due to saturated or repetitious coverage), or too dissimilar (e.g. entirely unrelated or disconnected from current editorial interests) \cite{Diakopoulos:2010jx}, though we defer such final judgment to the expert journalists in our evaluation. Alternative methods like topic modeling, clustering, or adjusting percentiles, can also be used, but our proposed method provides a straightforward and reliable approach with minimal parameters.

\subsection {Computing News Angles}
\label{subsection-methodsangles}

To generate news angles, we used GPT-3, a Transformer-based large language model (LLM) trained on 45 Tera-bytes of filtered English language data \cite{brown_language_2020}. An important issue is that text generated by GPT-3 is not based on factual knowledge or truth, but on making associations between consecutive words as they appear in the training data. This means GPT-3 can generate false statements, "hallucinate" information not in the prompt, encode problematic biases, and even copy text directly from the training data \cite{Bender2021OnTD, ji_survey_2022, lin_truthfulqa_2022}. Although we deploy the news angles in a human-in-the-loop system, this issue motivated us to do additional evaluation on the angles generated. 

To generate angles we provide a \textit{prompt} to GPT-3, in the form of the title and the abstract of the article concatenated together, and an instruction directing it to generate text. We first conducted several informal tests to identify techniques for generating news angles, and explored the impact of model fine-tuning and various model parameters on the accuracy and newsworthiness of news angles, as well as their uniqueness with respect to each other. These traits are important for the news angles to truly save time and cover multiple perspectives of journalistic interest. We also found that prefixing the abstract with the instruction statement, constraining it at the end to prevent continuation, and increasing frequency/presence penalty parameters generally increased output quality \cite{gero_design_2022, reynolds_prompt_2021}. We then conducted a small evaluation over a limited set of parameter-instruction combinations using a development dataset of 25 abstracts and a test dataset of 25 abstracts, each sampled randomly from the arXiv dataset of articles published in 2022. The following model parameters were varied in this evaluation: prompt, temperature, frequency penalty, and presence penalty\footnote{More detail in OpenAI's API Documentation: https://beta.openai.com/docs/api-reference/completions/create}. 

Adhering to best practices for the evaluation of such natural language generation (NLG) tasks, we conducted an expert-focused, intrinsic human evaluation where the first and second authors answered a series of questions rating GPT-3 responses for different criteria \cite{gatt_survey_2018}. Human ratings are expensive to collect, but are considered a better measure of the subjective quality of the text than automated metrics \cite{van_der_lee_best_2019}. For each news angle generated for an arXiv CS article, we rated it on a 5-point Likert scale for three criteria that we deemed suitable for this task, and which were inspired by similar criteria in the prior literature: fluency, accuracy, and news angle quality \cite{howcroft_twenty_2020}. Appendix \ref{appendixB} describes the definitions and scales used for each criterion. Prior work in NLG also emphasizes the effectiveness of such multi-item Likert scales where ratings for individual Likert items are ultimately aggregated through summation to denote the overall quality of the generated text \cite{amidei_use_2019, van_der_lee_best_2019}. We followed this practice to calculate an \textit{overall quality} metric as well.

To ensure the raters understood the definitions similarly, we piloted our definitions and iterated on the rating task until we could consistently apply them on the development set. We then rated three different news angles for each of the three specified criteria, for each of fifteen abstracts in the test dataset. We computed intra-class correlation coefficients (ICC) on the ratings for the \textit{overall quality} criterion obtained by averaging individual Likert ratings, to measure consistency between the two raters\footnote{Using the irr package \cite{gamer_irr_2019}} \cite{landis1977measurement, novikova2018rankme}. %ICC has been used to assess inter-observer reliability in similar NLG tasks \cite{} and ranges from 0 (no reliability) to 1 (perfect reliability).
We found that the ratings exhibited an ICC of 0.894 (n=15, p < 0.01), which indicates high inter-observer reliability. Following this, the second author rated the rest of the abstracts in the test set. We found the best instruction-parameter configuration, by comparing the average ratings for fluency, accuracy, news angle quality, and \textit{overall quality} across all abstracts in the test set\footnote{\textbf{Model}: text-davinci-002. \textbf{Prompt}: List three newsworthy headlines for this abstract: <text of title and abstract>. \textbf{Temperature}: 0.85 \textbf{Frequency Penalty}: 0.85 \textbf{Presence Penalty}: 0.85}.

\subsection{Designing the User Interface}
\label{subsection-uidesign}

\begin{figure}[t]
  \centering
  \captionsetup{justification=centering,margin=0.15\linewidth}
  \includegraphics[width={0.8\linewidth}]{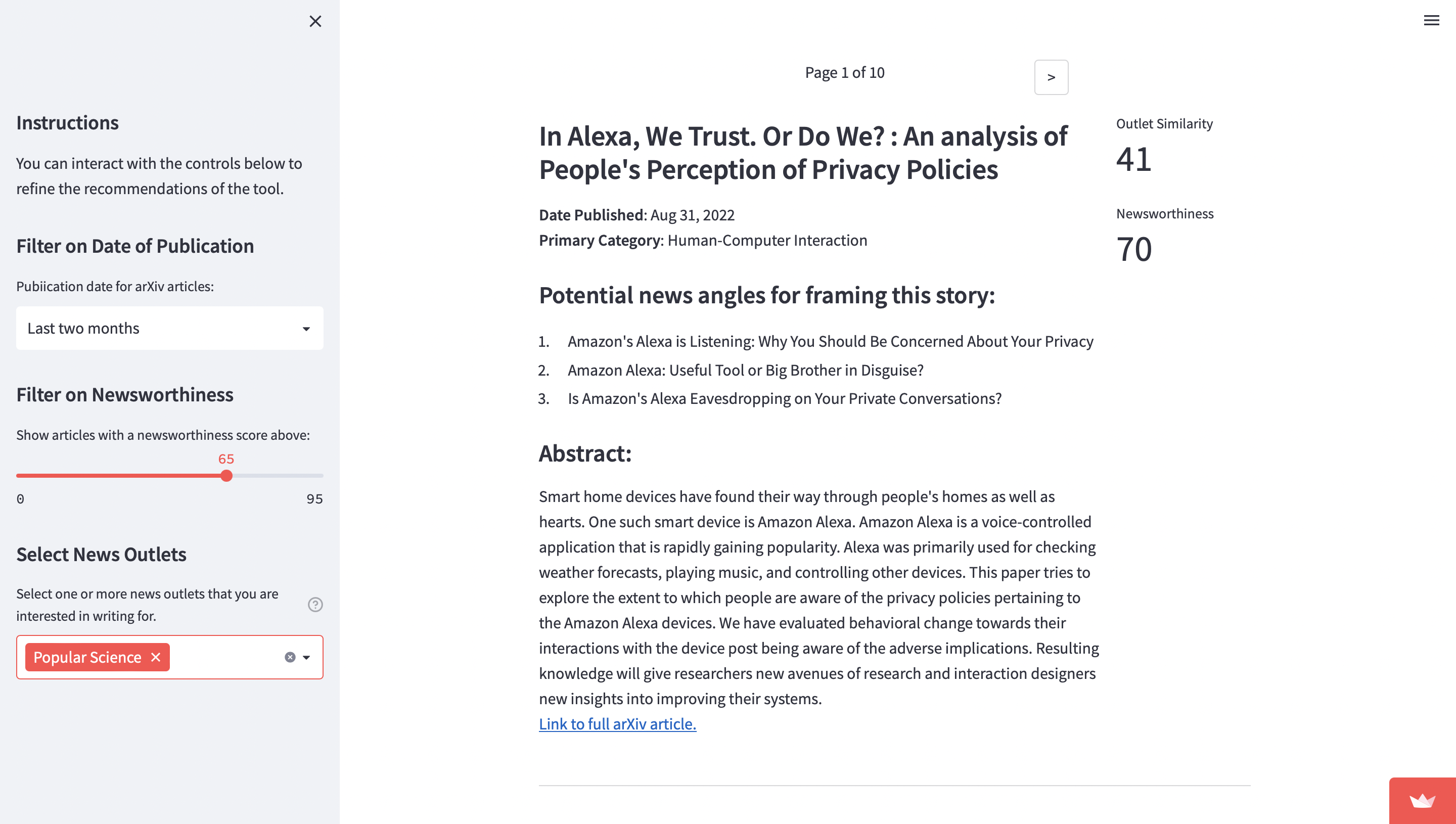}
  \caption{A snapshot of the user interface of the tool showing the sidebar with filtering options and an arXiv article that satisfies the criteria.}
  \Description[User interface for lead discovery tool]{A sidebar with different options and an article with its metadata, news angles, newsworthiness, outlet relevance}
  \label{fig:streamlit_screenshot}
\end{figure}

We created a browser-based tool\footnote{Using the Streamlit application} to enable the identification of newsworthy research from arXiv (See Figure \ref{fig:streamlit_screenshot}). The visual and interaction design of the tool's UI is largely drawn from recent scholarship at the crossroads of human-centered AI and journalistic endeavors \cite{broussard2019artificial}, including from previous CND tool research \cite{diakopoulos2021towards}. The initial view of the interface explains the data collection process and the computational information subsidies. It also presents a diagrammatic explanation of how to use the sidebar to filter and rank the articles. The sidebar offers controls to: (1) filter articles by date of publications, (2) rank or filter articles by \textit{newsworthiness score}, and (3) rank articles by \textit{outlet relevance scores} to specific outlets they selected. In case journalists selected multiple outlets, the relevance scores were averaged across them. Such nuances about the controls were provided in the sidebar on hover. 

Once participants selected their filters from the sidebar, the UI displayed all articles that fit the criteria in a paginated format, and ranked by either the \textit{newsworthiness score} or the \textit{outlet relevance score} depending on the user's choice. For each article, it displayed only the most relevant metadata that we believed could further support lead evaluation, and the assessment of the computational information subsidies. This consisted of the title, the date of publication, the abstract, and a link to the full article. Additionally, it displayed the \textit{news angles}, the \textit{newsworthiness score}, and the \textit{outlet relevance score}. News angles were displayed above the article's abstract, so that participants could quickly scan them before the more technically worded abstract. The link to the full article was provided below the abstract, as an avenue to further verification (e.g. full paper, authors, citations, etc.) beyond the context already provided in the UI. 

\section{Exploratory Evaluation Study}
\label{section-deployment}

\subsection{Participant Recruitment}

We recruited participants via non-probability sampling, specifically purposive sampling \cite{MartnezMesa2016SamplingHT}, by inviting science and technology journalists using online channels such as science journalists' listservs, posts on Twitter, and targeted outreach via email and Twitter messaging. Our screening procedure resulted in eleven participants who were freelancers or staff reporters for general or science-focused outlets. Six participants identified as female or nonbinary, and five participants identified as male. The participants covered different fields within science and technology, with a few tangentially covering computer science in their broader writing. This is important because our tool ranks articles from the Computer Science subject area and where it intertwines with other fields e.g. medicine, social science, etc. However, all participants did have experience with evaluating, writing, or pitching stories with technology-related angles, and with sourcing from preprints. This, combined with their general expertise in science journalism adds credibility and some degree of ecological validity to this work. Information about the varied professional contexts of the participants is provided in Table \ref{table-participants}. Participants were compensated for their participation with a \$50 Amazon gift card. 

\subsection{Participant Interviews}

All recruited journalists participated in an online scenario-based, semi-structured interview which was audio-recorded for later analysis (Min: 49 mins, Max: 71 mins, Median: 57 mins). After participants consented, the interview was started with some pre-usage questions to familiarize the interviewer with participants' current approaches to news discovery. Participants were then directed to the arXiv news discovery tool and a basic overview of its functionality was provided both visually and verbally. Once the participant had understood the tool's main controls, we invited them to begin using it under the scenario that they were looking for leads that they would want to further investigate for a news story. 

As journalists used the tool, the interviewer asked questions about their perception of the metrics provided, the controls they used, and the quality of the ranked articles. We also asked questions about how the displayed \textit{newsworthiness score} and \textit{outlet relevance score} aligned with their own perceptions of how newsworthy the lead was, or how well it fit their audience. If users asked questions about the system or the subsidies, their questions were answered by the experimenter. We sought feedback on how current subsidies or related new features could better support their work. We also asked about the manner in and extent to which journalists would prefer to engage with CND tools in their everyday practice. We further asked a series of post-usage questions to understand if the tool could mesh with or complement extant sources and practices for news discovery in their work. All interviews were conducted by the first author to maintain similarity in the broader lines of inquiry that were pursued. Appendix \ref{appendixC} provides a list of questions that were created to roughly guide these semi-structured interviews.

\begin{table}
    \scriptsize
    \caption{Participant Characteristics}
    \Description[Participant Characteristics]{A table with a list of participants and their roles, type of employment, and news outlets they write for}
    \label{table-participants}
    \begin{tabulary}{\linewidth}{LLLL}
        \toprule
        \textbf{Participant ID} & \textbf{Role} & \textbf{Type of Reporting} & \textbf{Type of Outlets Working for/Pitching to} \\ 
        \midrule
        P1 & Science and Technology Reporter & Staff Reporter/Freelancer & General news and Science-focused outlets \\ 
        P2 & Science Reporter & Staff Reporter & General news outlet \\ 
        P3 & Science Reporter & Freelancer & Science-focused outlet \\ 
        P4 & Science Reporter & Staff Reporter/Freelancer & General news and Science-focused outlets \\ 
        P5 & Science and Technology Reporter & Freelancer & General news and Science-focused outlets \\
        P6 & Data and Technology Reporter & Staff Reporter & General news outlet \\
        P7 & Science and Technology Reporter & Freelancer & General news and Science-focused outlets \\
        P8 & Science Reporter & Freelancer & Science-focused outlets \\
        P9 & Science Reporter & Staff Reporter & Science-focused outlet \\
        P10 & Science Reporter & Freelancer & General news and Science-focused outlets \\
        P11 & Science and Technology Reporter & Staff Reporter & Science-focused outlet \\
        \bottomrule
    \end{tabulary} 
\end{table}

\subsection{Thematic Analysis}

Interview recordings were transcribed and contextualized with notes to indicate when participants interacted with specific items in the UI. We conducted inductive, semantic thematic analysis \cite{Braun2006UsingTA}, and first identified important quotes pertaining to how participants saw the opportunities, problems, or tensions with computational subsidies, and imagined improvements or their impacts. We assigned these to open codes. We continually compared and contrasted these codes, while adding memos to record how they could be grouped at a more abstract level based on practices, constraints, and tensions in science journalism that can impact and or be impacted by computational subsidies. We iterated on this process until clusters of codes began to emerge and thematic saturation was achieved. We then axially coded the data to develop specific themes and sub-themes, and new memos were added to synthesize and summarize the themes. The co-authors discussed the open and axial codes through these iterations to refine, differentiate and group the emergent themes. The next section describes these themes in detail, providing examples and quotes for context.

\section{Findings}
\label{sec-findings}

Based on our thematic analysis of participant interviews we identified three main themes. The first theme highlights the \textbf{common dimensions and benefits of agency} experienced or desired by participants as they engaged with the computational subsidies, despite their diverse preferences and backgrounds. The second theme clusters the \textbf{distinct reporting contexts} that participants operated in (See Table \ref{table-participants}), and how these shaped participants' expectations of and interactions with the computational subsidies. The third theme delves into the tensions and conflicts of participants' \textbf{professional roles and societal responsibilities}, and examines how the introduction of computational subsidies may impact these in practice. Together, these themes highlight the many ways in which computational subsidies can impact science journalists' extant practices, whether it is at the level of influencing their individual information behaviors, addressing contextual information needs, or changing the news production process at a more macroscopic level.

\subsection{Common Dimensions and Benefits of Agency with Computational Subsidies}
\label{sec-findings-common}

The participants in our study had very diverse interests, experiences, and backgrounds. However, we identified some common patterns in how they utilized the subsidies. Specifically, the subsidies enabled participants to exercise more agency over discovering and making sense of individual articles, as well as over the broader corpus. Additionally, the participants interrogated the subsidies and expressed a desire for more information to further this agency and its benefits. Here we discuss these shared experiences and expectations from the participants, with an equal emphasis on the subsidies that elicited these responses. 

\subsubsection{Ranking and Filtering Help to Strategize and Guide Attention}

Participants across different reporting contexts saw the newsworthiness and outlet relevance scores as akin to traditional information subsidies that could help them strategize their news discovery process and "elevate certain papers" (P11). Participants tended to mainly explore and consider leads that were ranked highly by either of these scores. They applied different minimum thresholds for \textit{newsworthiness scores}, to filter out "articles that I would waste my time on" (P4) and obtain a minimal set of leads for examination. We received generally positive feedback for the effectiveness of the \textit{newsworthiness scores} in identifying relevant leads. Further, all the staff journalists that we interviewed pointed out that items with high \textit{outlet relevance scores} were at least topically aligned with how they perceived the selected outlet’s coverage. 

For the filtered leads, the precise values or minute differences in their scores were inconsequential in comparison to their relative rankings: "The exact similarity score never matters, for all the articles that show up on the first page, I will check them out equally" (P2). 

In cases where participants specifically navigated to the low-ranked leads, this actually engendered trust in the system’s rankings. As P6 noted after one such exploration with leads ranked very low for the \textit{newsworthiness score}:

\begin{quote}
"These seem to be very like in the weeds machine learning stuff … Makes me a little bit more confident when I go back to the super high newsworthiness stuff … This story [low newsworthiness item] probably is about the nitty-gritty specifics of how an AI works … And that's probably a good filter for my life. I like it."
\end{quote}

The continuous scores for newsworthiness and outlet relevance, and the ability to rank and navigate across leads via these scores thus allowed participants to filter leads and direct their attention more strategically, like in the case of traditional information subsidies. The ability to examine low-ranked leads, as opposed to entirely filtering them out or providing categorical classifications, also enhanced insights into and trust towards the algorithmic rankings. 

\subsubsection{News Angles Support Framing and Interpretation}

Participants saw the \textit{news angles} as a reliable way to identify hooks in a research article, and felt that sometimes the angles could "with a few changes, be made into a title or a pitch line" (P3). They also used \textit{news angles} to pinpoint the stakeholders affected by a scientific discovery. This allowed them to identify potential interviewees or the target audience for their stories. For instance, P1 was successfully able to use the \textit{news angles} on several articles to uncover the human characters in those stories: "I would start with talking to these end-users, who do these algorithms affect?" (P1). P6 used news angles to list potential interviewees: "On reading I can immediately think of what my contact list is going to start looking like … And I can interview them and really add some human character to this story." (P6). News angles occasionally provided this advantage even when the title and the summary were very obtusely worded, which is a widespread feature of scientific articles: 

\begin{quote}
 "Like this one … the title doesn't interest me one bit … none of this makes sense to me … So I wouldn't even begin to know how to kind of pitch it. But then the third bullet point, "A case study of COVID-19 projections in the US reveals race-based differences" … and like, oh, this is about COVID 19 and race-based! That's actually useful." (P8)
 \end{quote}

Participants also employed news angles to \textit{interpret} the research, sometimes reading them in lieu of the abstract: "This is much better than the abstract - which is what researchers write for their peers. There is a lot of jargon and terminology … News angles are summaries for non-experts like journalists" (P2). Another participant noted that this was "ultimately an efficiency gain … It helps me decide whether I even want to read the abstract, especially if they [the news angles] seem like they fit what I can see running in a certain publication". (P3) Thus, the news angles helped to understand the article and identify outlet relevance quickly, before the abstract had even been read. Yet, participants were cognizant that they still had to vet any interesting news angles for accuracy. They also made design recommendations that could further support the conversion of abstracts into an easily digestible format such as the automatic completion of abbreviations or providing a dictionary meaning if the mouse hovered over any jargon.

Participants also noted that the news angles were topical and easy to understand. In select cases, the angles were either very similar to each other or the title/abstract. This hindered their utility for finding hooks and understanding the article. Some users thus requested functionality to discard such redundant angles.

\subsubsection{Subsidies Can Enable Agency over Wider Sources}

Participants noted that traditional information subsidies such as press releases failed to provide enough user control over scientific sources, created a new information overload, and even restricted the range of sources they could efficiently explore. This theme emphasizes how the generalizability and user-friendliness of computational subsidies were perceived by participants as a solution that could mitigate this lack of control and source diversity.

Participants often relied on press release alerts to discover newsworthy science, which saved them the time of reading the full article, but also presented a new deluge of information and the possibility of false hype: "It can be really challenging when you receive like fifty press releases a day … then in this case for filtering I depend mainly on the title, which can be sort of misleading in many cases" (P7) Anecdotes of such misleading press kits from high-profile sources were also mentioned by other participants: "You know if there hadn't been a [Nature] press conference on this paper [about reviving pig organs posthumously], would it have gotten the same amount of coverage? Probably not." (P11). 

A second drawback of relying solely on press releases or high-prestige journals was reduced awareness of "a lot of newsworthy stories of science that …  gets into some society journal that doesn't have a press release engine" (P8). This reduced the source diversity of what participants could cover. Consequently, they also mentioned looking through preprint servers, lower impact factor journals, science newsletters, Google alerts and scientific blogs for newsworthy story ideas, but those presented their own information overload, since they "don’t have a lot of filtering options … there is no sorting … I have to scroll for each one of them and be like, okay, which one will be the most popular one?" (P4).

In contrast, participants felt that the computational subsidies and ranking mechanisms could enable them to easily and directly interface with scientific material from a variety of sources, and in their own time. One could "basically layer [the tool] on top of whatever repository of papers I am interested in … especially these places where science happens  [preprint servers, conferences, low IF journals], but it's like impossible to stay on top of" (P8). 

Participants also expressed that the simplicity of the subsidies and filtering process was less overwhelming than journal and preprint websites: "you don't want [a user interface] to be too busy and there to be too much information because then you scare people away like they look at it and there's too many choices … I don't want to interact with that at all … It's a good feature of this that it's very simple and plain" (P5). Some even considered using these subsidies for navigating publications that were accompanied by press releases, with P11 expressing hope that "Maybe machines will be better at it [than press releases]".  

Put together, expanded source diversity and greater agency also had a secondary benefit beyond the discovery of individual leads. Participants also saw these subsidies as a way to "get a very quick sense of the latest newsworthy papers" (P2) from various sources. Participants across different reporting contexts expressed this kind of interest in exploring the scientific breakthroughs without a specific story in mind, but in a manner structured by \textit{newsworthiness} or \textit{outlet relevance} rankings. Some  found that even brainstorming about a highly ranked lead without following up on it was beneficial. This process could help them file away scientific rabbit holes worth exploring when they had more time, or identify credible sources to interview for future stories. Thus, the subsidies not only enabled greater choice and agency for news discovery in the near-term, but also supported journalists in longer-term activities.

\subsubsection{Explanation and Transparency Can Enable Trust, Education, Verification}

While the ability to quickly "sort of browse through and then start fiddling [with sliders and controls]" (P6) gave journalists an initial idea of how the subsidies worked, they also sought explanations for the displayed \textit{newsworthiness} and \textit{outlet relevance scores}: "What does it [the newsworthiness score] mean? What went into it?" (P4). Participants gave examples of formats of explanations that could be helpful (e.g., scores for individual news values), and discussed how such interrogation of the tool could enable trust, debugging and education, and even expand horizons for storytelling. 

In cases where journalistic judgment conflicted with the tool e.g. if the tool reported a high newsworthiness but the journalist didn’t agree, explanations could help them determine "if there is something that I’m missing … to my educational benefit" (P10) or that "maybe I’m wrong about that … maybe my instinct is incorrect" (P5). One participant noted that explanation could also help to clarify whether they agreed with the score, but perhaps for different reasons. For instance, the journalist could appreciate an item’s high impact but if the tool said it was also high in controversy, it could then cue them to approach the lead in a new way. This was not implemented in the current UI but was requested by participants after they examined individual quantitative scores for leads. 

Participants' level of knowledge about how machine learning models worked impacted their expectations of transparency and explanations about the training dataset, computed subsidies, and ranking mechanisms. For instance, journalists with stronger intuitions about predictive models were interested in more technically sophisticated disclosures about the models, datasets, and training processes than others (e.g., inquiring about score averaging procedures). At the same time, technical intuition also created more reasonable expectations about the kind of transparency that was realistically possible. 

Relatedly, while most participants were familiar with GPT-3 and engaged in some degree of vetting of angles for accuracy, one participant specifically requested a list of the known caveats of this model. They emphasized that such transparency could empower journalists who are less familiar with GPT to make informed decisions about what they may need to verify or dig deeper into, without creating undue alarm. They noted: "If your journalists were hallucinating, you’d be having a problem. But when it comes to something like this with just news angles and, if you know while reading it, that there might be a little … wonkiness to it. It’s totally fine." (P6). 

\subsection{Context-Specific Experiences and Expectations of Computational Subsidies}
\label{sec-findings-context}

In this theme we explore how the diverse reporting contexts such as qualifications, experiences, and working styles of our participants (see Table \ref{table-participants}) affected their interactions with computational subsidies. We specifically delve into the following factors: journalists' audience focus, labor status, preferred writing format, and level of experience.

\subsubsection{Audience Focus: Generalist versus Science-Focused Outlets}

Whether participants wrote stories for general news outlets or for more niche, science-focused ones was a key differentiating factor in how they engaged with the presented subsidies. Participants writing for general news outlets focused on the recency of stories after ranking by \textit{newsworthiness scores}: "In a general newsroom … if it [a research article] was published a few days ago, it’s already old news." (P2) In contrast, journalists writing for science-focused outlets had a wider window on the date of publication due to differing priorities: "Some research is not relevant to me because it's new. It's relevant to me because it's an important finding. I would even have gone for the last two years." (P5). 

P1, P2, P4 and P7 who all worked at or pitched to the science desks of general news outlets, also looked for signals and suggested features that could help them understand what the popular reception for a story could be like e.g. whether the authors on an article were local to their region, what the social media traction was like, whether there were prestigious industry collaborators, etc. 

\subsubsection{Labor Status: Staff Journalists versus Freelancers}

Participants' usage and perceptions of the presented subsidies were also contingent on whether they were staff journalists at a single news outlet, or if they were freelancers who pitched stories to several outlets. 

For instance, both freelancers and staff journalists saw the \textit{outlet relevance score} as a crucial feature: "it doesn’t matter how newsworthy [a lead] is, if my outlet doesn’t cover that type of article, they won’t take it" (P4). However, this subsidy had a greater time-saving impact for freelancers, who would have to conduct this assessment for several outlets, not just one: "It takes time for a journalist to do their homework and check the media outlet so that they know if they should pitch this idea to this specific outlet or not. So I would say this score is very nice." (P7). 

Freelancers made further recommendations for how the outlet relevance could better save time and enable them to design pitches for many outlets in parallel: "I look for stories that I can pitch to Bloomberg, The New York Times, Scientific American, Wired, six or seven different outlets, just by angling the pitch a little differently each time. I guess if I’m going to spend 20 hours working on a pitch, I don’t want it to be just for one outlet." (P5) As a result, several freelancers were interested in not only the \textit{outlet relevance score} that was averaged across the outlets of their interest, but also a breakdown of the average score for individual outlets. This could let them see, for a given lead, which outlets and how many outlets it could be a good fit for.

\subsubsection{Writing Format: Study Stories versus Features}

Here we explore the patterns in engagement with subsidies based on whether participants wrote stories about single studies or experiments ("study stories"), or longer stories about broader patterns and trends in science ("feature stories").

Participants who focused on study stories appreciated that the information subsidies, rankings, and scrollable results could help them find tangible, newsworthy leads, often in response to scenarios that demanded speed: "I have an hour until my editorial meeting and I don't have any pitches … I could see this coming in really handy on time crunches like that." (P6) Participants also suggested the UI include interesting graphics from the ranked articles to improve its suitability to find study stories, since good graphics can be a selling point for editors within this format (P2, P8).

In contrast, participants who wrote feature or long-form stories mainly saw the provided subsidies as a way to discover and navigate a relevant set of secondary sources for stories they were already working on: "So I'm doing a story about AI. And if I just put that into Google, I'm going to get so many returns and I'm not going to get this newsworthiness score. I'm not going to get the potential news angles." (P5). These participants also suggested that implementing certain novel features could help them source new story ideas for long-form writing through subsidies. These features included (i) grouping similar articles that are published together, and (ii) suggesting articles that cited articles that were recently covered in the news. Such features could enable the use of "a single study as jumping-off point" (P11) for identifying novel trends and emerging fields.

\subsubsection{Experience: Level of Experience with Reporting}

Relatively inexperienced journalists had very different experiences and expectations with the subsidies in comparison to the seasoned reporters. 

P2, P5, and P11, who all had more than a decade of experience, spoke to having a well-tuned sense of what is newsworthy, and especially being able to infer that for their respective editors and audiences. These reporters paid limited attention to the newsworthiness score when initially presented with it: "I have not been looking at that [the newsworthiness score] just because … I feel like I have a very good sense after all these years of knowing what’s a story that I can sell." (P5) Given their experience, they also reported having easy access to a network of scientists and sources who could help contextualize information for them as and where necessary, thus serving as the journalists' own "mini peer-review committees" (P11).

In contrast, P4 and P10, who were both newer to the profession and also freelanced, both reported encountering trouble while discerning "what I find interesting versus what other people are going to find interesting" (P10), and both reported receiving recent rejections for pitches on account of their work being too niche. Thus, they found educational value in the provided scores, especially the \textit{outlet relevance score}. They also laid a heavy stress on potential features that could suggest suitable outlets and appropriate background context for a given lead e.g. suggestions of best outlet fit for an article, links to similar historical coverage from news outlets, Altmetric-style metrics of traction, author affiliations to check credibility, etc. This could enable a better understanding of outlet fit, and even help them to craft a convincing pitch for specific publications down the line. 

\subsection{Professional and Societal Responsibilities Impacted by Computational Subsidies}
\label{sec-findings-tension}

Participants shared insights on several conflicts and tensions that they must navigate in their practice, which stem from their professional and societal responsibilities.  These tensions relate to key issues such as exclusivity of leads, editorial and journalistic interests, and algorithmic curation in science communication. We discuss participants' reflections on how computational subsidies can impact these tensions and the various actors that are involved in them. 

\subsubsection{Exclusivity and Wider Subsidy Use}

Participants believed that the computational subsidies could provide them first-mover advantage on newsworthy leads. The study's inherent focus on preprints instead of embargoed research provided valuable exclusivity: "I am interested in finding original stories that no one else has picked up … preprints provide a little bit of an edge in terms of originality of the story" (P3). Generally, the subsidies were seen as a way to "monitor a preprint before it goes live" (P4) i.e. to reach out to interviewees and conduct background research before it was published in a peer-reviewed journal or under embargo. Thus, the subsidies could allow journalists to create original and exclusive stories, which they could verify in their own time, without rushing and risking mistakes.

However, some participants considered how wider use of these subsidies would diminish the exclusivity of the leads they viewed. They made assumptions about how their peers or competitors would use the subsidies, and how that would affect their own use: "If me and the 20 or 30 other journalists working at this level on this subject are all using the same tool, we're all seeing the same information presented. So how do I make sure that I get this story and not him or her? " (P6)  Another participant voiced a similar concern about suggested news angles: "If you provide these three angles for me and you’re providing also these three angles to other journalists, it means that every one of us will assume that the other journalists will use this, so it will end up that everyone is avoiding it" (P7). One even suggested an ability to do a "negative search" (P5) on popular topics to be able to explore exclusive and diverse leads.

Yet other participants alleviated these concerns via the subsidies themselves. For instance, P1 and P6 both navigated the UI to find a sweet spot for the \textit{newsworthiness score} that wasn’t too low or too high", so as to find the "hidden gems that others have missed, or could be given a new spin, or mixed with some other information" (P1). Another participant thought the homogeneity of news angles could even beneficially level the playing field: "To readily see the importance of an obscure paper, where the implications of that headline are not obvious to an average person. It's a skill that a lot of journalists have that gives them an advantage over other journalists. But [the news angles] can have a bit of a leveling effect on that." (P9)

These findings highlight a complex relationship between how subsidies can grant individual journalists more time to investigate diverse leads, but could also reduce uptake if journalists are concerned about a lack of exclusivity. 

\subsubsection{Editorial versus Personal Interests}

Journalists are public intellectuals and science communicators, but they are also creative workers who prefer covering certain themes and ideas more than others. This means that our participants often had a very specific idea of the topic, fields or subject areas they were looking to explore, from which they then sought out specific stories. For staff journalists, this was related to the specific beat they covered, whereas for freelancers, the topic revolved around whatever events had caught their attention in the news, or had emerged as a "side tangent" (P8) from previous reporting, or if "an editor puts a call for pitches on Twitter" (P10). Consequently, almost all journalists requested the inclusion of a search feature or a category-based filter, to be able to narrow the options down to the topics they were already interested in. 

Freelancers especially suggested more sophisticated personalization capabilities: "if there could be a way of putting my own work into it … like having the tool look at what I like, what I’ve written and match it up " (P3). Another participant also requested for the tool to learn his preferences based on feedback, so that he could "build a profile, how I define newsworthiness, for my point of view" (P7). Some journalists also had preferences for specific kinds of news angles that they preferred. For instance, one participant exhibited a strong interest in human-interest stories that explored "technology as a tool rather than the core of the thing" (P1). Yet another participant suggested personalization based on a journalist's self-expressed reading level and domain expertise. 

% full P1 quote: "technology as a tool rather than the core of the thing … how technology helps people with their lives" (P1)

Freelancers made most of these suggestions because they consider it vital to efficiently discover stories that interest them, as not all their pitches are accepted by editors, despite their effort: 

\begin{quote}
    "Finding those unique stories and then landing them, convincing an editor. Those are definitely the hardest parts … And that’s all before you ever have any agreement to make any money, because as a freelancer, you kind of have to balance your time … sometimes someone hands you on a silver platter and says ‘Write this.’ … And it’s definitely much easier and more lucrative, but rarely as intellectually satisfying." (P8)
\end{quote}

\subsubsection{Long-term Credibility of Algorithmic Curation}

The participants' use of computational subsidies was aimed at discovering leads that met their professional priority of producing newsworthy stories while fulfilling their overarching responsibility of creating stories that benefit society. Ensuring positive outcomes at both macro (societal) and micro (individual/professional) levels was crucial for the long-term credibility and success of the subsidies.

To ensure credibility at the the micro-level, some participants were looking for occasional rewards from the tool i.e. a useful lead once in a while: “The level of trust [with this tool] that’s required is actually pretty low … I’m going to vet it [a lead he pointed to] afterwards … but the thing that I’d need to know to continually use this is that occasionally, it will pop up something that is newsworthy for me … and that would suffice … if I got two stories a year off of this, I would check it weekly, at least.” (P9) Others were more likely to use the tool if they saw a continuous reward i.e. an improvement in its suggestions over time: “I am expecting the AI will not provide the best results in the first few times … But if I spend time to adjust my input and to give feedback, and if I can see that this enhances the results … then I would know it would save me time.” (P7) Building credibility over the longer-term is thus a nuanced and even potentially a personalized process.

At the macro-level, journalists were concerned with how computational subsidies could shape their and their audiences' attention to information. Notably, most of them expressed that human and algorithmic curators already played a tremendous role in their news discovery process, and were "obviously totally skewed as well" (P8). A tool like this was just one more curator among a slew of others that they drew from, albeit it sourced leads from an archive different to traditional subsidies. Journalists were still in control and required to exercise their judgment:

\begin{quote}
"[This tool] is similar to when you have to sift through press releases. And you know, just because it's a press release, or just because the news is embargoed doesn't mean that it's a story. So yeah, I do think that you would have to take it with a grain of salt … in this case, you know an AI program is generating that score in the case of EurekAlert or Nature, somebody at the journal or somebody at an institution is saying, "We think this is newsworthy". And it's still your job as the journalist to decide what to cover and what not to cover." (P11)
\end{quote}

Some especially viewed the machine-learned score rankings as a new proxy of quality separate from press releases, journal impact factors or Twitter threads, stating that they were "differently biased" (P8), since "a lot of preprints now get covered … only because some journalist has a source who emailed it to them, who is the PI or knows the researchers behind [it]. [This tool] at least does have some level of curation different from that" (P9). Some participants were however concerned about how this tool could detract societal attention from scientific stories ranked low for newsworthiness or outlet relevance. Using the score ranking feature, they investigated these stories explicitly to better understand "what are the things that are going to be lost in that low newsworthiness" (P6).  Tangible controls like navigability over machine-learned scores thus built some measure of credibility that the tool could support journalists' social responsibilities over the longer-term.

\section{Discussion}

As AI technologies become more prevalent in newsrooms, recent scholarship in HCI has called for more intelligible tools that incorporate professional values within domain-specific settings, and allow for users to effectively use, oversee, and adapt them \cite{jonesAIEverywhereNowhere2022, komatsuAIShouldEmbody2020, abdul_trends_2018, diakopoulosComputationalNewsDiscovery2020}. Thus, in this study, we aimed to uncover the various interactions, opportunities, and tensions that can arise when journalists in the specific domain of science reporting engage with CND tools. To achieve this, we conducted semi-structured interviews with journalists where they interacted with multiple prototype computational subsidies in a tool. The design of the tool's UI and the subsidies drew from prior work in centering user needs in interactions with AI-infused systems, and adapted it to the specific case of science journalists interacting with complex technical information. We then examined the patterns that emerged when science journalists used the subsidies, imagined adapting subsidies in their contexts, and considered CND's impacts on their values and practices. 

In the following subsections, we discuss the various interplays between science journalists’ practices and CND tools that their interviews highlighted, contextualize them with respect to prior work that we reviewed, and consider the design opportunities that might arise from our findings. In particular, based on our findings, we elaborate understandings of and implications for journalistic practices in light of the computational information subsidies we studied (Section \ref{disc-supportingPractices}), detail contextual and personal understandings of newsworthiness and news discovery (Section \ref{disc-newsworthiness}), and also step back to suggest future opportunities and features for CND tools (Section \ref{disc-features}). 

\subsection{Supporting Journalistic Practices}
\label{disc-supportingPractices}

In the following subsections we examine how the prototype computational subsidies and UI controls might support journalistic practices in terms of enabling agency in journalistic decision-making and in developing longer-term reliance on CND tools. 

\subsubsection{Agency in the Context of Journalistic Decision-making}

In Section \ref{subsection-sciencejournalism}, we described increased workloads and reduced resources that journalists dealt with as they engaged in news discovery from various channels. This can create new time pressures and lead to stress, burnout, and even reduced quality of reporting \cite{andersonViewTrenchesInterviews2023, powellWhatScienceJournalism2015a, powellWhatScienceJournalism2015b}. Not only did our participants echo some of these challenges, but they actively considered the potential of the prototyped subsidies to counter these issues in their own workflows. They described the information overload and bias posed by existing information subsidies like PRs, described the difficulty of navigating science articles from search engines and journal websites, and generally felt that the prototype subsidies and simple UI aligned with their informational needs and bolstered their agency when evaluating newsworthiness of leads within this complex ecosystem. 

We observed this editorial alignment and enhanced agency in a few different ways (Section \ref{sec-findings-common}). For the immediate task of discovering newsworthy stories, journalists not only benefited from the ability to navigate preprints, but even considered overlaying the subsidies on top of search engine results and publication websites to navigate those. This could allow them to scan and explore various sources while being guided by specific editorial values i.e. via \textit{newsworthiness} or \textit{outlet relevance scores} as operationalized in science journalism, rather than metrics such as view counts for articles or chronological order of publication. The \textit{outlet relevance} specifically allowed journalists to find stories that might align with editorial requirements of outlets they worked for or pitched to. Prior work has found this to be a factor that would support lead uptake if implemented \cite{diakopoulosComputationalNewsDiscovery2020}, and in this work we presented one potential operationalization of this criterion, validated its utility with a small sample of reporters, and even uncovered specific reporting contexts (e.g. freelancing) wherein this support could be central. Other ways of calculating this subsidy this could also improve its efficacy. Beyond this, news angles helped highlight specific newsworthy stories even when jargon-heavy titles and abstracts did not elicit immediate interest. Recent work in generating news angles using LLMs for political press releases similarly found that the technique helped journalists engage with complex material more quickly \cite{petridisAngleKindlingSupportingJournalistic2023}. Unlike this work however, we prompted the system to deliver diverse news angles by manipulating finer prompt wording and parameter settings. Providing angles along specific news values such as \textit{controversy} or \textit{novelty} as in \cite{petridisAngleKindlingSupportingJournalistic2023} could help users filter articles by their exhibited news angles - a functionality some participants expressly requested. 

In terms of the specific controls offered over the corpus of preprints and the subsidies, users benefited from the flexibility to select the ranking and filtering strategies themselves i.e. ranking based on either the \textit{newsworthiness} or the \textit{outlet relevance}, depending on whether they were freelancers or staffers. This exemplifies a specific case of how allowing users to choose and influence ranking strategies can improve their experience \cite{jannachUserControlRecommender2017}. Similar work in computational journalism \cite{parkSupportingCommentModerators2016} allowed comment moderators to adjust a ranking or choose ranking presets to help evaluate news comment quality. Our findings here broadly align with and express support for such a ranking based design approach that facilitates ranking choice or adaptation for different editorial purposes. We further found that support for ranking, filtering, and navigating by scores instead of providing outright recommendations helped participants streamline news discovery and exercise their own judgment, without making them feel like it constrained their autonomy. Prior work has also leveraged these benefits of ranking approaches \cite{diakopoulos2021towards, wangJournalisticSourceDiscovery2021, diakopoulosComputationalNewsDiscovery2020}. Beyond what has already been observed, we find a specific way in which this approach allows journalists to exercise their judgment: using the rankings, several journalists not only examined highly ranked items, but intentionally searched for hidden gems i.e. more exclusive leads from the lower ranked items. 

These insights may also hold for the design of CND tools to enable journalistic agency in contexts beyond science reporting, although adapting such a system to other beats would entail changes to how subsidies for ranking and filtering are computed. This could mean selecting domain-specific news values to operationalize for the \textit{newsworthiness score} (e.g. \textit{surprise} for sports news to capture newsworthy upsets), and even possibly optimizing for different criteria when prompting and evaluating LLMs (e.g. lifestyle news exhibits a different style of writing as compared to science reporting).

\subsubsection{Towards Reliance in the Longer-term}

Another way in which participants’ conversations around enhanced agency and editorial alignment took shape was in the way of discussing their longer term benefits (Section \ref{sec-findings-common}). For one, users found that the quick filtration and navigation could help with what prior work has identified as “backgrounding” \cite{diakopoulos2021towards} i.e. gaining broader perspective on scientific developments and trends. This can help journalists to build their news judgment in the longer term \cite{blum_field_2006} and improve their “nose for news”. Providing specific features to further support this backgrounding effort, such as greater historical context like provided to scientists by services such as Semantic Scholar, while calibrating it for the needs and time constraints of non-technical experts like journalists could be useful.

Journalists also suggested new features like local explanations for quantitative scores. This would improve their ability to act in an informed manner, aligning with prior calls for explainable interfaces in computational journalism tools \cite{jonesAIEverywhereNowhere2022}. We uncovered users' specific motivations for wanting explainability, along with their ideas for its implementation. This can inform the design of tangible features in future work. For instance, we found that while freedom to explore the interface built some level of initial trust, score explanations would better support trust over extended use, especially when users disagreed with scores and wanted to understand why. Disagreement could reflect either the tool being incorrect or present a learning opportunity to the journalist who might have missed something. Agreement with scores also led to a desire for explanation: so that they could understand if they had picked up on the same aspect of newsworthiness as the computed score. If not, that could broaden their reporting as well. This idea of building trust is thus also linked to two related functions that explanations have empirically been found to support in other domains: recognizing buggy system outputs, and learning from system outputs \cite{jannachExplanationsUserControl2019}. Another potential impact of explanations that our participants did not voice but which arises in the context of journalists' learning and growth is that they could specifically allow users to reflect on the values emphasized in their work over the longer term \cite{jonesAIEverywhereNowhere2022}. Actively designing explanations to better support these functions is vital, and even building on it by allowing users to report and dismiss scores and angles that are low-quality or even harmful would be needed, as recent scholarship around algorithmic contestability in HCI has suggested \cite{lyonsConceptualisingContestabilityPerspectives2021,
vaccaroContestabilityAlgorithmicSystems2019}. 

Journalists also desired transparency about training data and models, especially to reveal fallacies in generative models \cite{Bender2021OnTD, hongPlanningNaturalLanguage2021}. However, the extent of transparency sought broadly depended on the journalist's own technical expertise. Recent work similarly shows that effective explanations for computational tools in the news should generally match users' expertise \cite{nigatuCoDesigningTransparencyLessons2023} and promote easy understanding \cite{heuerExplanatoryGapAlgorithmic2021}. Conducting an evaluation of CND explanations and transparency based on technical expertise can help nuance these insights as well. Implementing internal audits of these systems could also provide a systematic understanding of what the computed subsidies could be missing, and bring to light broader bias patterns that arise if they are used uncritically \cite{rajiClosingAIAccountability2020}, leading to better design and well-informed usage. We also hope to explore more specialized audit methods for biases in LLMs in future work, such as red-teaming and human-AI collaborative auditing \cite{rastogi_supporting_2023, ganguli2022red}. Operationalizing different normative ideals in journalism - not just news values, but even values such as diversity, objectivity, etc. \cite{linOneThingJournalistic2022, komatsuAIShouldEmbody2020, deuzeWhatJournalismProfessional2005} - and understanding how subsidies might impact those for smaller samples could be a promising direction of work. 

Finally, our participants were also conscious of how CND tools can alter the functioning of the newsroom \cite{Pavlik2000TheIO}, and direct journalistic and societal attention at scale \cite{helberger_freedom_2020} (Section \ref{sec-findings-tension}). Some even used navigation and filtering to understand what kinds of stories were systematically elevated and which they might miss out on, thus conducting ad-hoc audits of their own, reminiscent of user-engaged auditing \cite{dengUnderstandingPracticesChallenges2023}. However, when describing the overall role that CND could play in their practice, they considered it to be analogous to other human and algorithmic curators in their news discovery repertoires, while also providing a novel signal of potentially newsworthy science. They voiced several pitfalls of relying solely on PR such as the limited sources and the potential for institutional bias, and saw CND as a way to add new perspectives in their reporting. They were not blind to its potential biases, but rather considered them in context of the indicators of newsworthiness they already relied on. The simple and navigable UI could this enable reporting diversity over time, by enabling access to sources lacking PR or impact factors \cite{thurman_human_2021}. It could also facilitate coverage across different kinds of article within these sources.

\subsection{Newsworthiness and News Discovery}
\label{disc-newsworthiness}

In the following subsections we elaborate our findings as they relate to understandings of newsworthiness, including its domain-specific but also contextual and personal dimensions, and specifically delving into the dimension of exclusivity. 

\subsubsection{Newsworthiness is Domain-specific, but also Contextual and Personal}

Our discussion so far has considered the idea of newsworthiness at a high-level of abstraction, and considered how CND tools can enhance agency and do so reliably as journalists evaluate newsworthiness. We now discuss the repercussions of our findings for what actually constitutes newsworthiness in our chosen domain, and how this might vary based on reporting context and personal interests. For context, in prototyping the subsidies, we drew from across journalism studies, CND and HCI (Section \ref{subsection-cnd}) to establish relevant news values, motivate evaluation of outlet fit, and generate news angles. We argued that these allow for the incorporation of broader editorial values, and observed that our participants found them useful.

However, we found newsworthiness is not only dependent on news values relevant to broader science news production, but also depends on reporters' own professional priorities and the demands of their roles (Section \ref{sec-findings-context}). For instance, certain news values outside the scope of the current evaluation were deemed relevant for different reporting contexts. Providing options to view them or toggle their incorporation into \textit{newsworthiness scores} could help. Reporters for generalist news outlets could be supported by news values that relate to the \textit{popular reception} of a story e.g. proximity to their specific geographical context, celebrity names, and share-worthiness \cite{badenschier_2012, harcupWhatNewsNews2017, trillingNewsworthinessShareworthinessHow2017}. Writers of study stories could also toggle an option to view interesting \textit{audio-visual material} from scientific articles, which are known to lend a special appeal to study stories \cite{badenschier_2012, harcupWhatNewsNews2017}. 

Further, our findings also highlight a need for more personalized conceptions of newsworthiness, especially according to the thematic or creative interests of journalists (Section \ref{sec-findings-tension}). We find that this could balance journalists' own interests with editorial assessments of newsworthiness or outlet fit for the stories they pitched. This was especially desired by freelance journalists, who often have some autonomy to seek out stories instead of being assigned them. They actively search for stories that align with their interests and are feasible for them to cover given their available resources. However, they also face the challenge of pitching stories to various types of outlets, necessitating a delicate balance between satisfying editorial interests and pursuing their preferred subjects. We remind the reader that with the structural and institutional transformations in the field of science journalism that have caused a large-scale shift to freelance work for science journalists \cite{dunwoody2014science}, it is vital that we design interventions that cater to the needs of this growing population. Issues around time compression that we have reviewed are also exacerbated for freelancers, since they must invest substantial effort not only in sourcing stories but also in bridging gaps in their scientific knowledge and managing their professional reputations without any institutional support \cite{andersonViewTrenchesInterviews2023}. 

Personalized rankings accompanied by the newsworthiness subsidies can help journalists make these judgements during the news discovery process. For instance, future tools could boost articles similar to stories journalists have written before, if they have a particularly narrowly-focused beat, or generate news angles similar to the types of angles present in the journalist’s prior work. Offering customizability on the specific news values aggregated within the \textit{newsworthiness score} as suggested above \cite{parkSupportingCommentModerators2016} could specifically support journalists who prioritize certain news values on account of either personal or the above mentioned contextual reasons. Participants even recommended personalization on the basis of their reading level, long-term usage patterns in the tool, and specific kinds of news angles they enjoyed covering (e.g. human interest). Recent work has found evidence in support of personalized, LLM-generated news angles as well, but this can require time and experience for phrasing and scoping prompts, which journalists may not always possess \cite{petridisAngleKindlingSupportingJournalistic2023}. We echo their call for future work that explores support for journalists writing and refining LLM prompts for news discovery workflows. Ultimately, we excluded implementations of search and personalization in this study to conduct a minimal evaluation of the subsidies. However, displaying explicitly personalized results alongside the existing information subsidies in longer-term deployments could help understand if this truly supports journalists in balancing their creative and personal preferences with organizational goals. 

Such evaluation is necessary because algorithmic personalization also presents trade-offs: while it caters to journalists' personal preferences for stories and topics, it can also raise ethical concerns around biases in how information is ranked and filtered \cite{diakopoulos2021towards, dehaanInvisibleFriendFoe2022}. While the \textit{newsworthiness score} and \textit{outlet relevance score} have some kind of journalistic and editorial values encoded into them in both our work and prior scholarship in CND \cite{diakopoulos2021towards, wangJournalisticSourceDiscovery2021, liu2017reuters, diakopoulosGeneratingLocationBasedNews2020}, personalization may not provide this benefit. Some of the questions our users raised around algorithmic curation and how it could detract attention from certain stories applies to personalization-based ranking as well. 

One way to navigate the trade-off between catering to journalists' interests while avoiding personalization that reduces the variety and quality of their reporting is to conduct internal audits of explicit personalization mechanisms, as suggested for computational subsidies above. Allowing users to navigate, sort, and filter articles by personalized relevance scores could also help them explore lower-ranked items more intentionally, similar to how they did for low \textit{newsworthiness score} items. This can go further in helping them to systematically look through items that are not recommended to them right off the bat, but in a way that is manageable and can be done in their own time. An additional benefit is also that this level of control could help build trust in the personalized rankings like it did for computational subsidies in the study and like prior work has shown \cite{harambamDesigningBetterTaking2019}. Studies of the benefits users derive, if any, from such control mechanisms over longer-term deployments are needed. Science writing is as much an endeavor in creativity and story-telling as it is in engaging with normative values in journalism to tell relevant stories. To be able to support the former without compromising on the latter would be important to navigate for end-users. 

\subsubsection{Discovering News Stories: Exclusivity, Collaboration, or Both?}

The emphasis laid by our participants on exclusivity (Section \ref{sec-findings-common}) shows how traditional news values, as operationalized within the \textit{newsworthiness score} in our case, paint an incomplete picture of newsworthiness; the exclusivity of a lead can often determine its final selection \cite{schultzJournalisticGutFeeling2007}. This desire for exclusivity within CND tools has been seen before \cite{diakopoulosComputationalNewsDiscovery2020}, but received a deeper consideration in our context from both freelancers and writers of study stories, neither of whom wanted to spend time pitching leads that would already be covered by competitors (Section \ref{sec-findings-tension}). Some were also concerned that such a common repository of computational subsidies could reduce perceived exclusivity and diminish the overall uptake of leads and news angles from CND tools. Some suggested features to enable exclusivity evaluation for leads, which might be supported using recent news coverage or aggregate viewing patterns within the tool. 

However, whether exclusivity should be incorporated into such tools is as much a judgment about journalists’ informational needs as it is about their normative values. Our findings hint at this duality as well: one participant weighed the homogeneity of news angles against their educational benefits for inexperienced reporters. A broader homogeneity in the news may also enhance robustness and consensus around scientific issues from the public's perspective, with different reporters bringing their vetting skills, unique ideas, and organizational values to a story. Such a philosophical shift away from exclusivity and toward sharing of resources historically led to the formation of the Associated Press in the U.S. It has also yielded returns in the context of investigative journalism in terms of open-sourcing of data, methods, and best practices which facilitate collaborations across newsrooms and produce more robust and reproducible journalism \cite{mullerGatekeeperGateopenerOpenSource2023}. Konow-Lund et al. even describe the extensive benefits of global collaboration in the context of investigative journalism, and how digital tools have mediated and supported these efforts \cite{konow-lundTransnationalCooperationJournalism2019}. 

One could conceptualize such an orientation within CND tools for science journalism as well, where signals of collective interest (e.g. aggregate viewing patterns for leads) could be framed as opportunities for collaboration, rather than invoking problematic tensions surrounding speed or urgency that can often threaten reporting quality \cite{powellWhatScienceJournalism2015b}. Two recent developments in the coverage of science and technology also support this potential shift: (1) increased time pressures that we discussed before mean that journalists might benefit from collaborating with peers to combine their efforts in the public interest, as well as to hone their own reporting skills (2) scientific collaboration and technological supply chains are increasingly globalized \cite{dongCenturyScienceGlobalization2017, grayGhostWorkHow2019, sarkarEnoughHumanAICollaboration2023}, meaning that collaborations between journalists across geographies could reveal wider insights about the research conducted by both universities and companies alike, and still lead to the production of distinct stories for all their individual audiences. At the same time, the time pressures and reducing pay rates already compel science journalists to hastily produce stories in order to stay competitive \cite{andersonViewTrenchesInterviews2023}, and so any efforts for designing tools for collaboration need to actively reduce the friction and time overhead of setup. Based on our findings we suggest that there is a potentially rich design space that can be explored in future work for thinking about how to balance exclusivity with potential for collaboration in CND tools to support journalists.

\subsection{Implications for CND Interfaces and Features}
\label{disc-features}

In the following subsections we describe what we think are some implications and also opportunities to further develop computational supports and subsidies to enable journalistic news discovery. 

\subsubsection{Contextual Information Needs Supported by Flexible Interfaces}

One clear opportunity emerging from our findings is for designing modular, flexible interfaces that can be configured according to journalists’ reporting contexts (Section \ref{sec-findings-context}). While prior work has suggested that configurability of tools and their interfaces can support their usability and integration into journalistic workflows \cite{diakopoulos2021towards, gutierrezlopezQuestionDesignStrategies2022}, our work here contributes vital nuance to this by identifying several different axes along which such configurability can be offered (e.g., level of experience, writing format). Here we describe some of the ways this can be achieved, via examples of features that users can toggle on or off. Offering them as intelligent defaults based on user contexts and backgrounds can also reduce the initial overhead of setup.

For instance, inexperienced reporters looking to learn about outlet fit and pitching might be supported by an option to view detailed context for quantitative \textit{outlet relevance scores}, e.g., popular or recent news stories from a chosen outlet that a lead is semantically similar to, or news articles based on prior work from the same scientists. Given their lack of an established network or contacts, they might even benefit from being suggested domain experts based on the similarity or newsworthiness of their research, and consent to being suggested. Feature writers could be provided options to cluster recent articles by similarity or sort similar articles chronologically, so that they could spot new trends in science. Writers of study stories would benefit from the seamless scrolling of leads similar to our UI. 

Reporters' varying time availability and workloads also motivate certain features to support contextual variation. Our participants conveyed that features such as bookmarks, scientists' contact information, and email alerts with customizable frequencies and thresholds for \textit{newsworthiness scores} could help them archive leads for future coverage, which aligns with previous successful implementations \cite{diakopoulos2021towards, wangJournalisticSourceDiscovery2021}. Although these features were not included in our study, which focused on evaluating needs and expectations for specific computational subsidies, their potential usefulness, particularly in longer-term deployments, is evident.

Ultimately, flexibility in tool design can have tangible advantages, as it encourages journalists to invest more effort in pursuing leads \cite{diakopoulosComputationalNewsDiscovery2020}. However, providing it at the different levels discussed in this section is also difficult because designing complex user interfaces and aggregating different information sources to align with diverse preferences can be conceptually and computationally challenging \cite{diakopoulosComputationalNewsDiscovery2020}. Future work concerning configurability for computational news discovery must explore how it can be embedded in a way that aligns with journalistic values, but is also technologically feasible. It must also delve into how configurable features interact with journalists' expectations and assumptions in a real environment, which can further impact how these tools are perceived and used \cite{Orlikowski1994TechnologicalFM, petre_all_2021}.

\subsubsection{Generative Support For Sense-making and Reporting}

Recent developments in generative text models have elicited great excitement within journalism, and our participants’ initial interactions with this technology was not so different: the generated news angles in our study enabled journalists to identify newsworthy narratives and elicited largely positive feedback (Section \ref{sec-findings-common}). Such a generative approach appears to align with the idea that newsworthiness is not necessarily inherent to events, but is rather generated by contextualizing and framing information \cite{Lester80}. Still, the positive response was somewhat surprising given the known concerns with model hallucination, the potential to output inaccurate text, and the lack of transparency around training data. However, the participants largely seemed nonplussed and willing to work with the generated angles as suggestions that they further validated. From a quality standpoint, there were still the occasional cases where the generated text significantly repeated the title and abstract, or didn't offer substantial variance amongst the set of angles generated. It's possible that these issues could be mitigated in future work by utilizing stricter values of parameters such as the frequency penalty, or by automatically filtering out generated angles in cases of high semantic similarity to the abstract or to other angles. Future implementations could develop different prompting strategies that could be keyed towards specific news values like conflict, negative consequences, as described above \cite{petridisAngleKindlingSupportingJournalistic2023}, so that presented news angles exhibit more deliberate variation. Either way, we learned that incorporating some notion of uniqueness into an evaluation protocol would be helpful, so that this dimension can be measured and minimized with respect to model parameters or prompts. 

A specific reason for participants’ positive responses was that the news angles could sometimes help them identify not only interesting narratives from the research, but also potential audiences, stakeholders, and interviewees for the resulting news stories. Prompting large language models that are augmented by knowledge bases to specifically generate these items for a given scientific abstract could thus provide further support for journalists’ ideation process \cite{lewisRetrievalAugmentedGenerationKnowledgeIntensive2021}. Prompting models to generate these responses based on the discussion and conclusion sections of articles could perhaps offer greater accuracy or relevance of the responses as well. Based on the prior discussions concerning personalization and outlet relevance, one can also imagine designing prompts to tailor news angles specifically for journalistic interest or for greater outlet fit. Generating text based on standard taxonomies of news angles could further permit the filtering of leads based on the news angles they exhibit \cite{motta_analysis_2020}. This would enable further agency and control over the subsidies in a way that produces personalized outcomes. If journalists were to participate in this process of tailoring the prompts, that also could enable the development of ways for educating them about using generative models. 

\subsection{Limitations}

In this study, we recruited eleven participants who operated in varied professional contexts and environments (e.g. staff journalists vs. freelancers, science features writers vs. study story writers, etc.). We believe this provided us with an adequate sample of interviewees who had wide-ranging experiences and approaches to their work. However, our sample was geographically limited to include participants who wrote for national-level news outlets in the United States, and further work is required to investigate how our findings generalize to other regional contexts. Our sample contained an almost equal split of male and female identifying participants, but it exhibited limited racial diversity. Demographic diversity could impact how journalists prioritize news values and the specific constraints they operate under, and future work should consider these factors when examining the perceptions and usage of computational news discovery tools. Finally, the lead discovery tool we developed solely covered Computer Science articles on arXiv, which limits its applicability in the wider field of science journalism. In future work we hope to extend it to other scientific domains, enhancing utility for the broader community of science and technology journalists.

\section{Conclusion}

In this work we uncover the potential opportunities and implications of CND tools for professional workflows and responsibilities in science journalism. To accomplish this, we prototype different computational subsidies and an interactive UI to create a tool that can support news discovery, and interview reporters as they use it to discover and brainstorm leads. We find that the presented subsidies are largely aligned to editorial values, and can help journalists exercise greater agency in their interactions with the vast landscape of complex scientific work. Beyond this, subsidies also offer a range of utility that is contingent on specific reporting contexts. We also uncover various trade-offs that journalists make within their practice that CND can sway, or even new tensions that CND itself can introduce. Based on these findings, we reflect on how CND can broadly support agency in journalistic practices and do so reliably over the longer term; allow for more personal, contextual and even peer-collaborative conceptions of newsworthiness in news discovery; and create novel functionality during news discovery via leveraging flexible interface design and generative models. Ultimately, CND tools designed  with an eye towards these nuances of their socio-technical context can not only streamline individual journalists' workflows, but could shift their dynamics with editors and peers, make journalistic work more sustainable in the longer term, and ultimately further the public interest by enabling more scientific information for broader audiences.

\begin{acks}
This work was funded by the National Science Foundation through Award No. IIS-1845460. We express our gratitude to the journalists who engaged in our interviews, offering both valuable insights and inspiration. We also extend our thanks to Darren Gergle, Ágnes Horvát, Ayse Hunt, and Vien Nguyen for their substantial support and constructive feedback during various stages of this project.
\end{acks}

\bibliographystyle{ACM-Reference-Format}
\bibliography{bibliography}

\newpage
\appendix

\section {Appendix A}
\label{appendixA}

We collected news articles published in the following news outlets, and computed \textit{outlet relevance scores} for all arXiv articles with respect to each.

\begin{table}[h]
    \caption{News Outlets for Calculating outlet relevance Scores}
    \Description[News outlets for similarity scores]{A table with the names of news outlets, URLs, and the type of work they publish}
    \label{table-outlets}
    \begin{tabulary}{\linewidth}{LLL}
        \toprule
        \textbf{Name} & \textbf{URL} & \textbf{Type of Outlet} \\ 
        \midrule
        ArsTechnica & https://arstechnica.com/ & Science/Tech News \\ 
        Futurism & https://futurism.com/ & Science/Tech News \\ 
        NewScientist & https://www.newscientist.com/ & Science/Tech News \\ 
        The New York Times & https://www.nytimes.com/ & General News \\ 
        Popular Science & https://www.popsci.com/ & Science/Tech News \\ 
        Popular Mechanics & https://www.popularmechanics.com/ & Science/Tech News \\ 
        Quartz & https://qz.com/ & General News \\ 
        Salon & https://www.salon.com/ & General News \\ 
        ScienMag & https://scienmag.com/ & Science/Tech News \\ 
        Scientific American & https://www.scientificamerican.com/ & Science/Tech News \\ 
        Stat & https://www.statnews.com/ & Science/Tech News \\ 
        TechCrunch & https://techcrunch.com/ & Tech News \\ 
        MIT Technology Review & https://www.technologyreview.com/ & Science/Tech News \\ 
        The Conversation & https://theconversation.com/us & General News \\ 
        VentureBeat & https://venturebeat.com/ & Tech News \\ 
        VICE & https://www.vice.com/en & General News \\ 
        Vox & https://www.vox.com/ & General News \\ 
        The Washington Post & https://www.washingtonpost.com/ & General News \\ 
        WIRED & https://www.wired.com/ & Science/Tech News \\ 
        The Verge & https://www.theverge.com/ & Science/Tech News \\ 
        \bottomrule
    \end{tabulary} 
\end{table}

%TC:ignore 
\newpage
\section {Appendix B}
\label{appendixB}

Each individual angle generated by GPT-3, for each abstract and each parameter combination, was rated on a five-point Likert scale from 1 (Strongly Disagree) to 5 (Strongly Agree) for \textit{fluency}, \textit{accuracy}, and \textit{angle quality} (See Table \ref{table-likert}). \textit{Overall quality} was obtained from the mean ratings of all criteria. 

\begin{table}[h]
    \caption{Evaluation Criteria for News Angle Generation}
    \Description[Criteria for evaluating news angle generation]{A table with the names of individual criteria and their definitions}
    \label{table-likert}
    \begin{tabulary}{\linewidth}{p{.2\linewidth}p{.7\linewidth}}
        \toprule
        \textbf{Criterion} & \textbf{Definition} \\ 
        \midrule
        Fluency & The text is fluent and readable, without grammatical errors  \cite{espinosaFurtherMetaEvaluationBroadCoverage2010, howcroft_twenty_2020}.\\ 
        Accuracy & The text accurately reflects the material presented in the scientific abstract  \cite{espinosaFurtherMetaEvaluationBroadCoverage2010, howcroft_twenty_2020}.\\ 
        Angle Quality & The text centers one or more news angles on the presented scientific abstract for journalists  \cite{opdahlOntologiesFindingJournalistic2021, motta_analysis_2020}.\\ 
        Overall Quality & Mean of fluency, accuracy, and angle quality. \\ 
        \bottomrule
    \end{tabulary} 
\end{table}

\newpage
\section{Appendix C}
\label{appendixC}

\textbf{Pre-Usage Questions}
\begin{itemize}
    \item How would you describe your work? Freelance? On assignment? A mix of both?
    \item What are the key sources where you come across or look for leads to report on? What are your experiences with arXiv?
    \item What is the easiest part of your lead discovery process as it exists now?
    \item What is the hardest or most frustrating part you face during your lead discovery process as it exists now?
\end{itemize}

\textbf{Feature: Newsworthiness Score}
\begin{itemize}
    \item What do you think about this metric of newsworthiness? How useful would it be?
    \item Does this metric align with your sense of what would be newsworthy in your own work? Why or why not?
    \item Was there a specific lead in the ranking that you thought was or was not newsworthy or at least that you might be curious to pursue a bit more?
    \item What is your level of trust in this measure of newsworthiness to uncover interesting leads?
    \item Are there any problems this newsworthiness score might present to how you judge potential leads?
\end{itemize}

\textbf{Feature: Outlet relevance Score}
\begin{itemize}
    \item What do you think about this personalization based on relevance to your targeted news outlets?
    \item Is the relevance of a potential lead to your targeted outlet a criterion in your work when you are selecting stories?
    \item Is this feature effective or distinguishable in your opinion?
\end{itemize}

\textbf{Feature: News Angles}
\begin{itemize}
    \item How do the provided news angles impact your decision-making process about a specific lead, if at all?
    \item Did any of the provided news angles provide creative sparks for what might make something interesting to report on further?
    \item What could make these news angles more effective to aid your decision-making process?
\end{itemize}

\textbf{Post-usage Questions}
\begin{itemize}
    \item Is there something missing from this tool that you expected based on the description?
    \item Are there any features that you think might enhance the utility of this tool for how you’d want to use it? Any further information we can provide to make this task easier?
    \item Can you see yourself using a recommender system with some or all of these features as part of your regular workflow?
    \item What might keep you from using this in your regular workflow?
    \item  Do you have an ideal tool or a bucket list for one that might help you uncover potentially interesting leads from the latest developments in science and technology?
\end{itemize}
%TC:endignore 

\end{document}